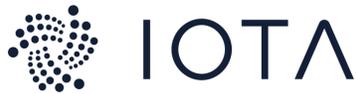

# Report on the energy consumption of the IOTA 2.0 prototype network (GoShimmer 0.8.3) under different testing scenarios

Louis Helmer, Andreas Penzkofer

IOTA Foundation
Pappelallee 78/79, 10437 Berlin, Germany

louis.helmer@iota.org
andreas.penzkofer@iota.org

# Abstract

The high energy consumption of proof of work-based distributed ledgers has become an important environmental concern. Bitcoin, for example, consumes as much energy in a year as a developed country. Alternative consensus mechanisms, such as proof of stake, have been shown to use drastically less energy than proof of work-based DLTs. For example, the IOTA DLT, built upon a directed acyclic graph (DAG) architecture, uses an alternative consensus mechanism that requires significantly less energy than other DLTs. Because the (DLT) space is constantly and rapidly evolving, the question of how much energy DLTs actually consume demands to be continuously studied and answered. Studying the energy consumption of alternative DLTs is important as it contributes to improving the understanding of the general public that not all cryptocurrencies use excessive energy resources. Previous research into the energy consumption of the IOTA network has shown that an optimization in the overall protocol correlates to an optimization in energy consumption. The planned IOTA 2.0 update, built upon the GoShimmer research prototype, promises to further optimize the protocol by removing the network's centralized Coordinator. This report presents the results of measuring the energy consumption of a private GoShimmer network while comparing these findings to previous research into the current mainnet, which is called Chrysalis. The main findings of this report are that the IOTA 2.0 research prototype shows both improvements and increase in the energy consumption metrics compared to the Chrysalis network. Additionally, this report defines a model to estimate the total annual energy consumption of an IOTA network. This model should be significant for future research as it enables a way to estimate the total cost of running the IOTA network as well as its carbon emissions. Moreover, having an annual power consumption metric allows for better objective comparisons to different DLTs.

# Table of contents

# 1. Introduction

The urgency of climate change causes us to reevaluate the energy consumption and efficiency of the products and services that underpin our daily lives. One example is the growing criticism of the energy consumed by proof of work (PoW)-based distributed ledger technology (DLT) projects, including Bitcoin[1] and (with the increased use of non-fungible tokens, or NFTs[2]) Ethereum[3]. It is appropriate, therefore, that cryptocurrencies should also be included in the energy consumption debate, especially if they aim to play a significant role in the future global monetary infrastructure.

Research from Alex de Vries for the Digiconomist Platform estimates that the Bitcoin network currently consumes around 204.50 TWh annually[4]. This can be compared to the total energy consumption of the country of South Africa, which consumes an estimated 202 TWh annually[5]. However, even though these figures are alarming, one should interpret this comparison as a means to make the energy consumption figures of Bitcoin tangible to the average person who is not familiar with electricity measurements. The limitation of this comparison is that one compares apples and oranges. Nevertheless, the question arises: why does Bitcoin as a monetary system need to consume such a large amount of energy? The answer: In order for Bitcoin nodes to add new transactions to the ledger, they must solve a cryptographic puzzle that requires a vast amount of computing resources. However, other consensus mechanisms, such as proof of stake (PoS), have shown that consensus can be achieved without the immense energy requirements by at least three orders of magnitude[6].

The IOTA 2.0 protocol introduces several new concepts, such as a novel consensus mechanism[7,8], and a new type of access control algorithm[9]. Through the latter, the protocol aims to address (together with other challenges) the energy consumption inefficiency by removing PoW. The lightweight design allows for low computational demand, meaning that it can run on low-powered devices (Raspberry Pis, for example). Many of the design decisions were influenced by IOTA's vision of enabling the software to run on a wide spectrum of IoT devices.

Previous research into the energy consumption of the current IOTA main network (called "Chrysalis") was published on May 14, 2021[10]. However, to protect the network against attacks, the Chrysalis network still uses a centralized node run by the IOTA Foundation (called "Coordinator") and uses a small PoW requirement as spam protection. While the Chrysalis network is already considered green in comparison with other DLTs (for the energy consumption of one Bitcoin transaction, one billion IOTA transactions can be sent)[11], the PoW still contributes significantly to the overall energy consumption of running the Chrysalis network. With the launch of the IOTA 2.0 prototype on June 2nd, 2021, a prototype software (called GoShimmer) was released and operated in a public testnet[12], in which the

---

Coordinator is removed. Although PoW is still employed in the current version to protect the network against spam attacks, a novel access control is being implemented, which allows it to phase out PoW. In the future, the Chrysalis network will be replaced by the Coordinator-free IOTA 2.0 network. With a new consensus mechanism, access control and improved network architecture, an additional reduction in the energy consumption of the IOTA 2.0 network in comparison to the Chrysalis network is expected.

The energy consumption of the GoShimmer network will be measured by running a private network of three nodes on a Raspberry Pi 4B. Measurements will be conducted under different testing scenarios. For the purpose of this study and to align this study closer to the case where the IOTA Congestion Control Algorithm (ICCA)[13], which controls the access, is fully implemented, we set the PoW in our experiment to a low value in the prototype.

The objective of this report is to study the energy consumption required to process a single message[14] and to determine the power consumption per GoShimmer node. As testing was done on the prototype network (0.8.3), the values presented in this report might change in the future based on improvements made to the 2.0 protocol. Future updates to this report will be published.

# 2. Literature review

As aforementioned, previous research into the energy consumption of the Chrysalis network has been published on May 14, 2021. The energy consumption analysis of the Chrysalis network was made in comparison to the legacy IOTA 1.0 network. This was done because the second phase of the Chrysalis upgrade introduced many performance improvements to the IOTA protocol. These changes reduced much of the computational load that nodes need to carry in order to run the network. The previous measurements were done in a private Chrysalis network, with the Coordinator node being deployed on a laptop, while two nodes were run on one Raspberry Pi 3B+ and one Raspberry Pi 4. A custom breadboard circuit was set up, with which the power consumption of each Raspberry Pi was measured with a Texas Instruments INA219 and leveraged with a breakout board by Adafruit. There were four tests. The first measured the energy consumption per transaction while spamming 50tps with remote PoW (done by the Coordinator node), the second spammed 100tps with remote PoW, the third spammed 0.0730 transactions per second while doing local PoW and the last measured the energy consumption of running a node at no spam (0tps). The reduction in energy consumption due to the Chrysalis upgrade (in different testing scenarios) is between 33% and 95%.

The author of the initial benchmarking remarks that one should resist the urge to make any assumptions regarding the linearity of the energy consumption/transaction. This is illustrated by the fact that the energy consumption per transaction of the 50 tps spam rate is higher than the energy consumption per transaction of the 100 tps spam rate. The author explains that this is due to transactions being issued much faster at a higher spam rate and, since energy is equal to power multiplied by time, less time issuing transactions results in less energy spent.

This report aims to expand upon the initial research made on the Chrysalis network and apply a similar testing set-up to determine the energy consumption metrics for GoShimmer.

---

[13] Cullen, A., Ferraro, P., Sanders, W., Vigneri, L., & Shorten, R. (2021, July 14). *Access control for distributed ledgers in the internet of things: A networking approach*. Retrieved February 07, 2022, from https://arxiv.org/abs/2005.07778
[14] A message is a data transaction

# 3. Methodology

The testing set-up to measure and determine the power consumption per GoShimmer node, as well as the energy consumption per message required at different spam rates, is shown in the table below. The results of this testing setup can be compared to the initial Chrysalis benchmarking.

| Test name | Explanation |
| --- | --- |
| Reference | Is the measurement of the total average power consumption of the Raspberry Pi 4B. Thus it is the base consumption of the device. No node software is running. |
| Resting | Is the measurement of the total average power consumption of a GoShimmer node (where the private network is running at a base activity of 2-3mps). No spam rate is active. |
| 50mps | Is the measurement of the total average power consumption of a GoShimmer node under a network load of 50mps. |
| 100mps | Is the measurement of the total average power consumption of a GoShimmer node under a network load of 100mps. |
| 200mps | Is the measurement of the total average power consumption of a GoShimmer node under a network load of 200mps. |

## 3.1. GoShimmer Private Network

The testing scenarios were conducted on a private GoShimmer network with three nodes (two peers and one faucet[15] node). The private network setup was achieved by utilizing the docker private network tool in the tools directory of the GoShimmer repository. See the *Software* section for further details.

## 3.2. Measurement

What was measured?
To measure the power consumption of the different testing scenarios, we will measure the number of joules (SI unit of energy) per amount of time (seconds). This will give us the power consumption in watts, which is joules per second.

How was power consumption measured?
To measure power consumption during the experiments, the USB-C energy meter *TC66C Ruideng* was employed. It was attached to the Raspberry Pi from which measurements were collected. The collected data was then sent to a device in CSV format. For a setup diagram and detailed explanation of how the meter works, please refer to the *Hardware* section.

---

[15] A node that is able to dispense tokens

Temperature

The room temperature at which measurements were made was around 22°C. The temperature of the Raspberry Pi 4B was around 60°C while running the docker private network.

## 3.3. Measurement accuracy

The volt and current measurement accuracy by the TC66C, as written in the manufacturing manual, are:

| Current | ± 0.1% |
|---|---|
| Voltage | ± 0.05% |
|  | Based on independent testing: ± 0.02% |

**Volt Accuracy of TC66C**
YouTube user "TheHWcave" published their research on the volt accuracy of the TC66C; they found that it was measured to stay within 0.02% above a voltage of ≈4V[16]. As the Raspberry Pi 4B requires a minimum 5.1V, we can determine that the voltage measurement had an accuracy range within less than ±0.02%.

*Figure 1 - Volt measurement accuracy of TC66C*

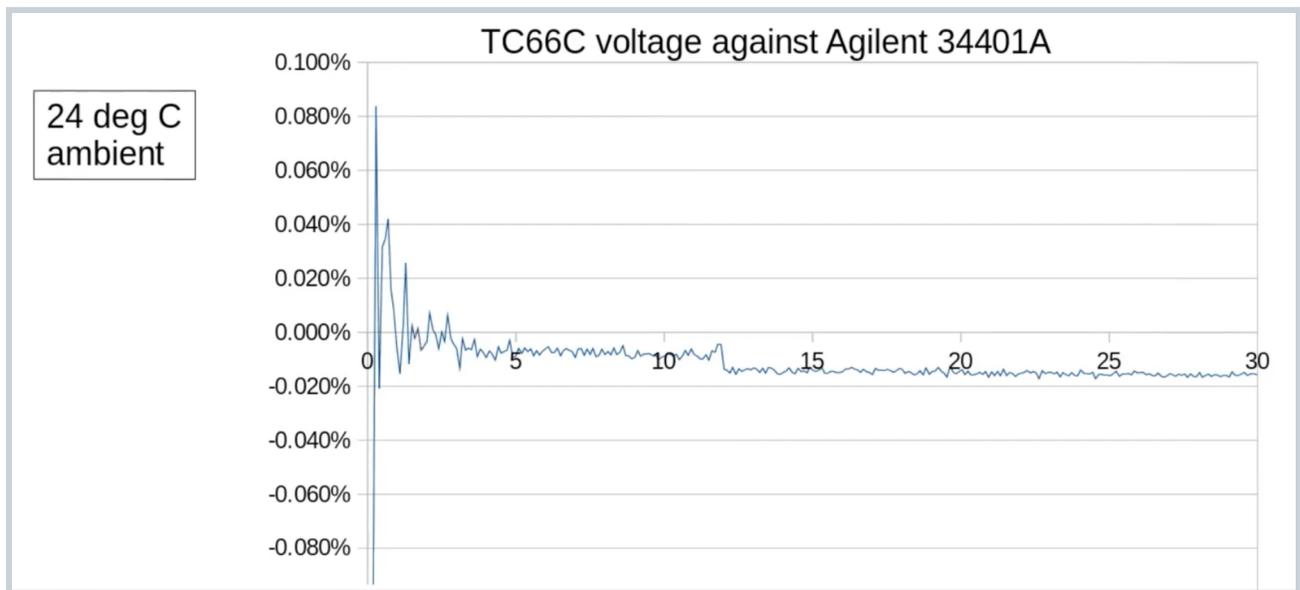

---

[16] TheHWcave. (2021, February 26). *Ruideng TC66 / TC66C USB Type-C Tester*. Video. Retrieved February 14, 2022, from https://www.youtube.com/watch?v=rOlhibDUJgs&t=557s

## 3.4. Hardware

The hardware used throughout the testing was:

| Running the private GoShimmer network | Measurement collection |
| --- | --- |
| <ul><li>Raspberry Pi 4 Model B Rev 1.1<ul><li>2GB RAM</li></ul></li><li>USB-C power supply</li><li>LAN-cable for ethernet connectivity</li></ul> | <ul><li>TC66C energy meter</li><li>Micro-USB power supply</li><li>Computer</li></ul> |

### 3.4.1. Setup diagram

Below is a visualization of the testing setup.

*Figure 2 - Testing setup diagram*

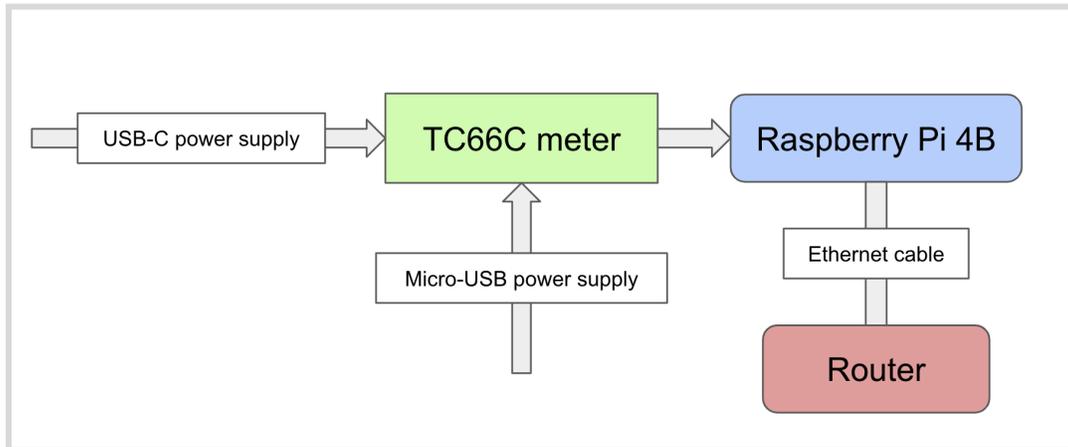

The TC66C meter was additionally supplied with power through a micro-USB cable to solely measure the power consumption of the Raspberry Pi 4B as the output reading.

### 3.4.2. Measurement device - TC6CC

Below is a simplified model of the TC66C meter created by *TheHWcave*[17]. All USB type-C signals directly pass through the meter. The differential signals are represented by the yellow lanes. The devices attached to both ends of the TC66C have no information that the TC66C is inserted between them. For the measurements, there is a 15milliOhm shunt resistor in VBUS to measure the current. Additionally, the INA226 Texas Instruments monitor is used to measure both a shunt voltage drop and the bus supply voltage. Note that if the power to the TC66C is coming from the socket end and it is not supplied with an external micro-USB cable, then the output reading will include the TC66C's own power consumption.

---

[17] TheHWcave. (2021, February 26). *Ruideng TC66 / TC66C USB Type-C Tester*. Video. Retrieved February 14, 2022, from https://www.youtube.com/watch?v=rOlhibDUJgs&t=557s

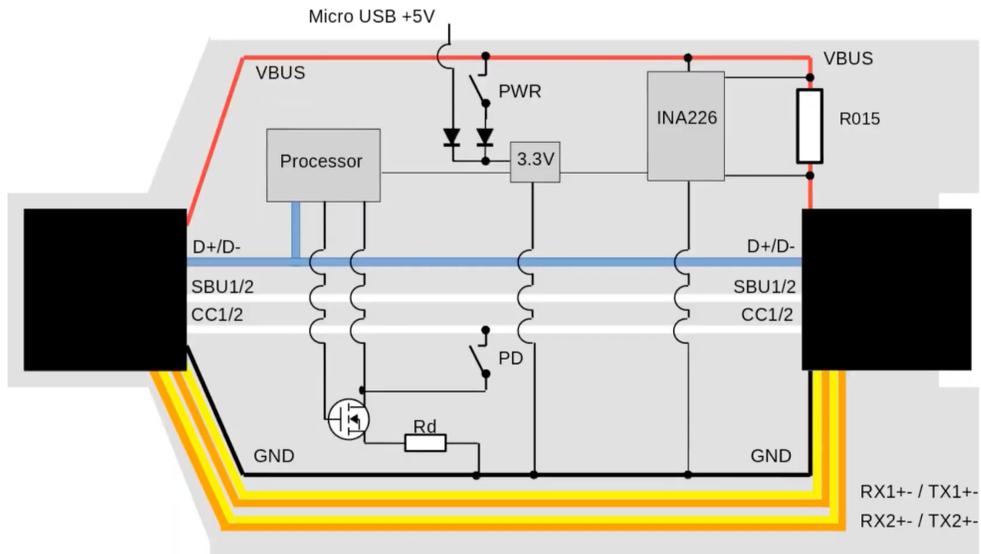

*Figure 3 - Simplified model of TC66C*

## 3.5. Software

### 3.5.1. Node software

The node software used for testing was the 0.8.3 version of GoShimmer[18].

### 3.5.2. Docker Private Network

The docker private network used for testing consisted of three nodes: two peer nodes and one faucet[19] node. The network was run with both the graphical analysis tool Grafana[20] and the distributed Random Number Generator (dRNG) committee disabled. All three nodes were run on one Raspberry Pi device. The dRNG was disabled because the progress of the dRNG is done by committed servers, which do not care about their energy consumption. Furthermore, the dRNG message is just a normal message from the point of view of a receiving node. As mentioned in the introduction, the reason why the PoW difficulty is set to two is to align this study closer to the case where the IOTA Congestion Control Algorithm is fully implemented.

---

[18] https://github.com/iotaledger/goshimmer.git
[19] A node that is able to dispense tokens
[20] https://wiki.iota.org/goshimmer/tutorials/monitoring

*Figure 4 - diagram of the private network setup*

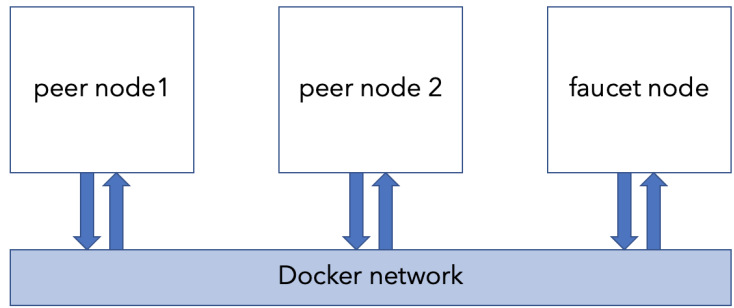

The configuration of the docker private network was as follows:

| PoW | Faucet |
| --- | --- |
| <ul><li>Difficulty = 2</li><li>NumThreads = 1</li><li>Timeout = 10 seconds</li><li>parentsRefreshInterval = 300ms</li></ul> | PoW difficulty = 12[21] |

**Linearity of the processing of messages in the utilized private network**

A metric that will be presented in the Results section of this report is the power consumption per GoShimmer node. As we are collecting combined measurements from a total of three nodes, the metric has to be calculated by dividing the measurements by three. However, this calculation assumes that all three nodes process the same amount of messages and occupy similar CPU and memory usage. Through collecting the analytics of each node, it can be concluded that we can indeed divide the average measurements linearly to get the metric of the energy average consumption per node. Below are two charts that show the CPU usage and memory usage for each node.

*Figure 5 - CPU usage chart - 9311 (blue), 9312 (red), 9313 (yellow)*

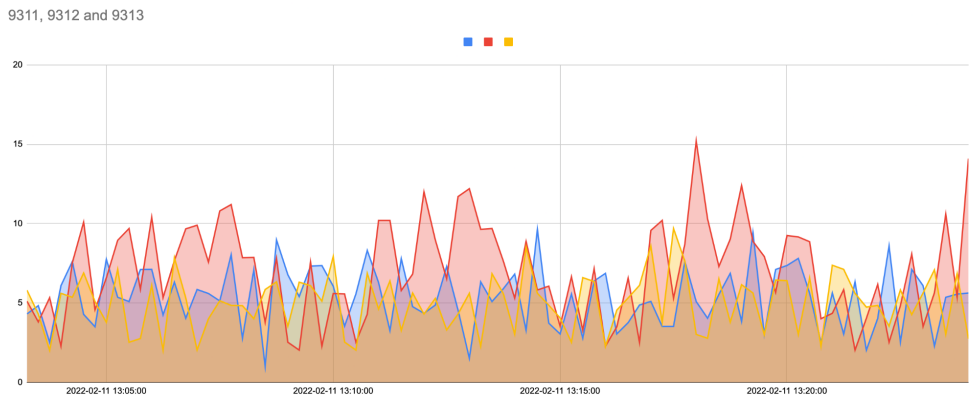

---

[21] The faucet PoW difficulty refers to the PoW necessary to request funds, which is an operation not used in this experiment.

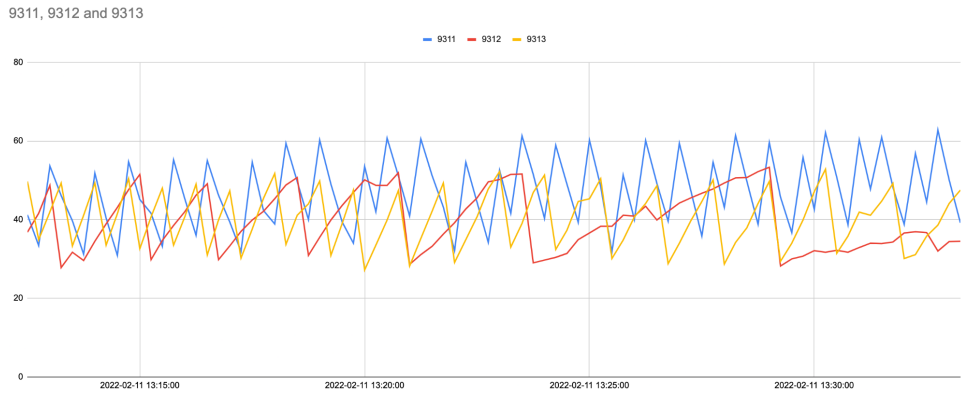

*Figure 6 - Memory usage chart - 9311 (blue), 9312 (red), 9313 (yellow)*

Moreover, as seen in the tables below, all nodes have on average an almost equal amount of messages stored in their databases, as well as the same amount of scheduled messages and a consistent tip pool of one. This means that all nodes in the network seem to process the same messages eventually. This demonstrates that we can divide the measurements by three. The measurements were made for durations of 30 minutes and separated from the power measurements.

| Node analytics | Node 1 (9311) | Node 2 (9312) | Node 3 (9313) |
| --- | --- | --- | --- |
| Messages in database | 4203 | 4207 | 4204 |
| Tip pool | 1 | 1 | 1 |
| Scheduled messages | 4540 | 4544 | 4541 |
| Average MPS rate | 1.5 | 1.5 | 1.5 |
| Average CPU usage | 5.35% | 7.07% | 5.01% |
| Average Memory usage [MB] | 47.08 | 39.69 | 40.37 |

Collecting the data above was made possible by enabling the Prometheus ports in the docker compose file of the docker private network. Prometheus and Grafana were run on a laptop (Prometheus was configured to read). The GoShimmer local metrics Grafana dashboard was utilized to visualize and collect the data of each node / instance in CSV format.

### 3.5.3. Spamming tool

The spam rate was set by utilizing the command line tool (curl). The commands were executed in the /goshimmer/tools/docker-network directory. Only data messages were spammed.

| Starting the spam at 50mps at an uniform rate. | curl --location 'http://localhost:8080/spammer?cmd=start&rate=50&imif=uniform&unit=mps' |
|---|---|
| Stopping the spam. | curl --location 'http://localhost:8080/spammer?cmd=stop' |
| Data message size | The spammed messages had a size of 127 bytes (123 bytes of message size and "spam" as a payload with 4 bytes) |

### 3.5.4. Raspberry Pi OS

The OS of the Raspberry Pi used in testing was the headless Lite 64bit version of the RaspiOS[22]. This OS was chosen to remove any unnecessary energy consumption from the equation and because a desktop interface was not needed for testing purposes.

### 3.5.5. Golang-Go

Throughout the setup process, the programming language *Go* had to be installed.

### 3.5.6. Docker Engine

| Client: Docker Engine - Community | Server: Docker Engine - Community |
|---|---|
| Version: 20.10.12<br>API version: 1.41<br>Go version: go1.16.12<br>Git commit: e91ed57<br>Experimental: true | Version: 20.10.12<br>API version: 1.41 (minimum version 1.12)<br>Go version: go1.16.12<br>Git commit: 459d0df<br>Experimental: false |

### 3.5.7. Docker Compose

docker-compose version 1.29.2

---

[22] Raspberry Pi Foundation, https://downloads.raspberrypi.org/raspios_lite_arm64/

# 4. Results

To make the results tangible for the average person, the results will be compared to the power consumption of a common light bulb in watts [W]. 60[W] means that the light bulb requires 60 joules per second [J/s] to function.

| Common light bulb power consumption | 60[W] |
|---|---|

The measurements made were current (ampere, A) and voltage (volts, V) over time (seconds) of the Raspberry Pi under the testing scenarios. Based on Ohm's law, the power consumption per second was then calculated.

> *Power [W] = Voltage [V] * Current [I]*

Given that the power consumption of the GoShimmer network fluctuates and the data in this report represents a sample, this report will also provide the minimum (lowest power consumption average recorded) and the maximum (highest power consumption average recorded) for the results of the different testing scenarios.

## 4.1. Unit conversion table

Throughout this report, we refer to different units of power and energy. The table below shows the conversion between the different units.

| | | | |
|---|---|---|---|
| Power | 1 Watt [W] | = | 1 Joule / second [J/s] |
| | | | 1000 milliWatt [mW] |
| | | | 0.001 kiloWatt [kW] |
| Energy | 1 Joule [J] | = | 1 Watt second [Ws] |
| | | | 0.00027 Watt Hour [Wh] |
| | | | 0.00000027 kiloWatt hour [kWh] |

## 4.2. Trials and measurement duration

The table below shows the duration of each test measurement and how many times a measurement was repeated. The 200mps test was conducted at half the trials and half the trial duration because the standard error already reached a sufficiently low value at 25 trials. Additionally, the 200mps spam rate was very volatile, which means that the measured result is not as accurate as the lower spam rate tests. Moreover, the mps rate for every second was not a collected data point, thus we could not specify the average mps rate for the average measurement of the 200mps trial. The 200mps results should thus rather be seen as an approximation of the power consumption at around 200mps. With a fluctuating spam rate, it is not meaningful to continue measurements if we cannot state to which mps rate the average result applies.

| **Methodology** | Reference | Resting | 50mps | 100mps | 200mps |
|---|---|---|---|---|---|
| Trials | 50 | 50 | 50 | 50 | 25 |
| Duration of one trial | 10 minutes | 10 minutes | 10 minutes | 10 minutes | 5 minutes |

## 4.3. Power consumption results

The graph below shows the two main energy consumption metrics (energy consumption per message and power consumption per node) calculated from the results presented in this report, in comparison to running a one-watt LED for one second. The energy per message metric is specifically for a network activity of 50mps because this metric changes based on the mps rate. The following pages will highlight this relationship. When looking at the graph, one can see that the energy consumption metrics of GoShimmer are very low when compared to running a LED. This is also highlighted by the fact that the bar representing energy consumption per message is barely visible. However, one should note that running a node and processing a message is only possible by running the GoShimmer software on a hardware device. Therefore this comparison includes the energy consumption of the hardware device used in testing.

*Figure 7 - GoShimmer energy consumption metrics compared to running a 1W LED for one second*

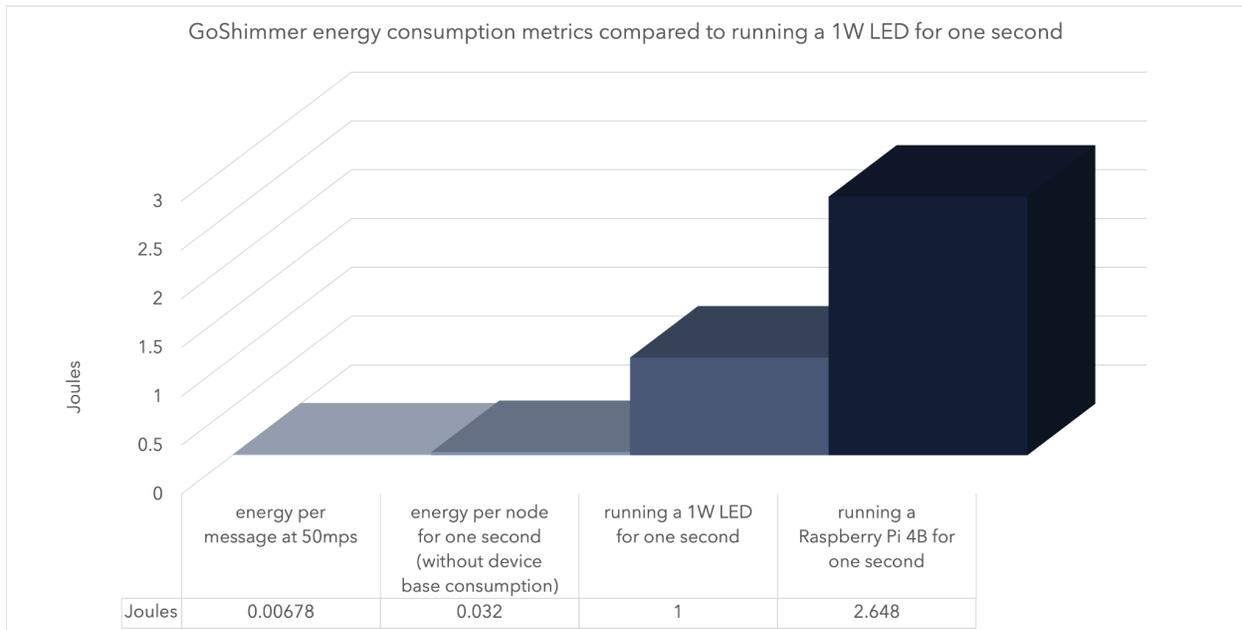

| | energy per message at 50mps | energy per node for one second (without device base consumption) | running a 1W LED for one second | running a Raspberry Pi 4B for one second |
|---|---|---|---|---|
| Joules | 0.00678 | 0.032 | 1 | 2.648 |

The graph below shows the average results of the measured total power consumption of the GoShimmer network under different tests. Tables 1, 4, 5, 6, 7 and 8 tabulate Figure 1.

*Figure 8 - Power consumption results - GoShimmer 0.8.3 on Raspberry Pi 4B (2GB RAM)*

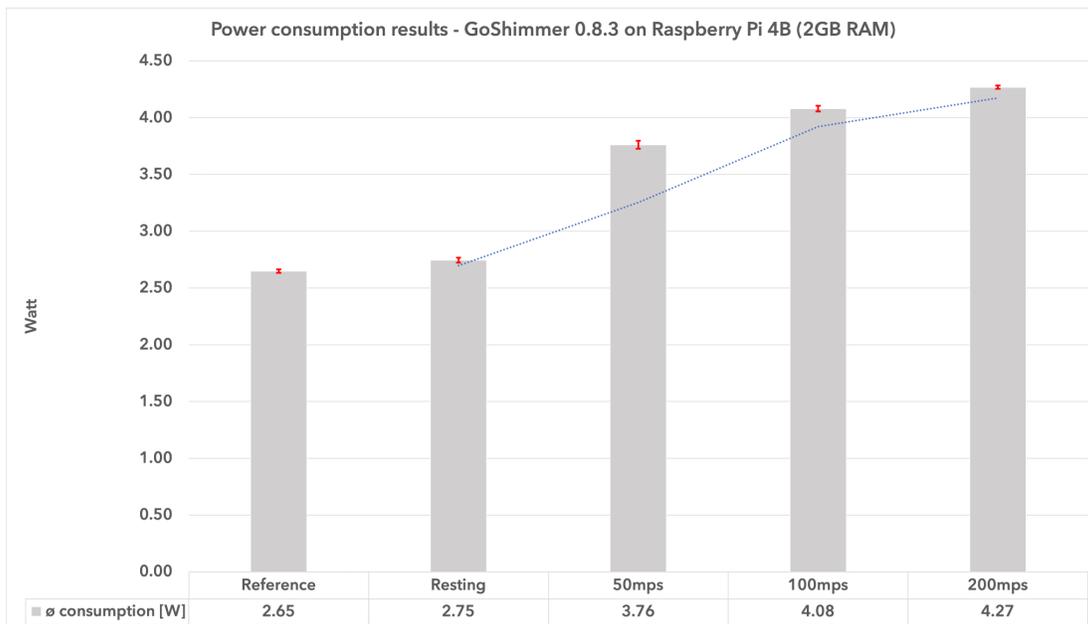

| Table 1 | Reference | Resting | 50mps | 100mps | 200mps |
|---|---|---|---|---|---|
| [W] | 2.648 | 2.745 | 3.761 | 4.080 | 4.268 |
| [mWh] for one second | 0.7355 | 0.7624 | 1.045 | 1.133 | 1.186 |

As shown above, the jump from Reference consumption to Resting consumption is very small. This tells us that running GoShimmer at base activity (with no active spam) requires little power. The majority of the power consumption comes from running the Raspberry Pi 4B itself. Moreover, the trend line and the values in the legend below suggest that when the spam rate is increased the power consumption does not increase linearly.

The graph below shows the results of the total average power consumption per GoShimmer node under different tests without the Reference (Raspberry Pi consumption). Tables 2 and 7 tabulate the data from Figure 2. The standard errors for this graph were divided by three.

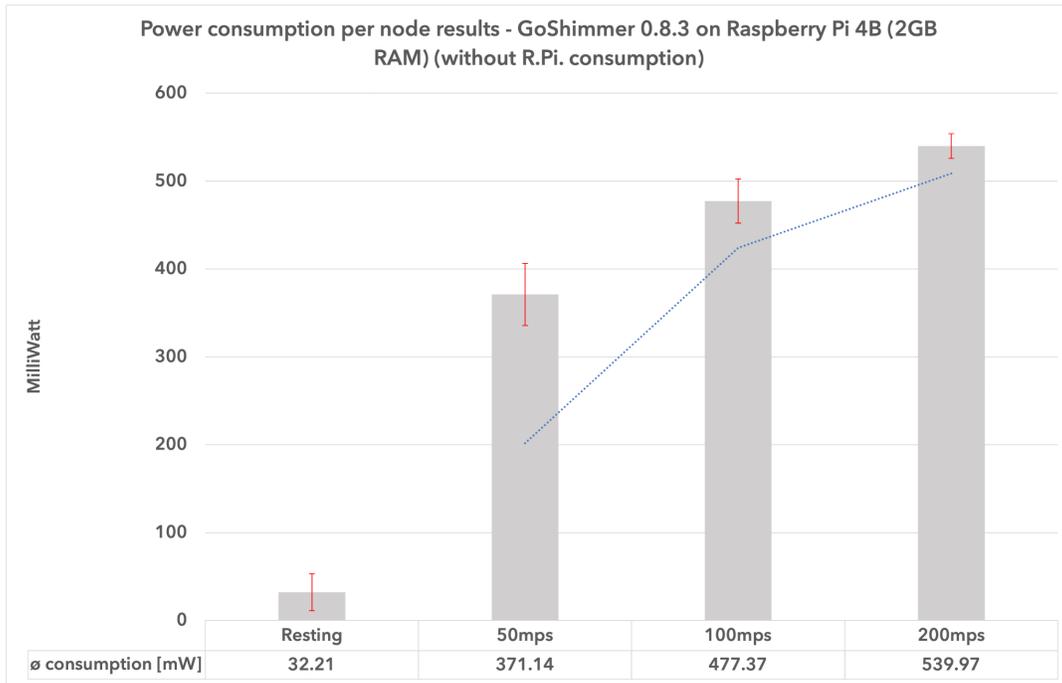

*Figure 9 - Power consumption per node results - GoShimmer 0.8.3 on Raspberry Pi 4B*

| Table 2 | Resting | 50mps | 100mps | 200mps |
| --- | --- | --- | --- | --- |
| [mW] | 32.21 | 371.14 | 477.37 | 539.97 |
| [mWh] for one second | 0.0089 | 0.1031 | 0.1326 | 0.1500 |

## 4.4. Energy per message consumption results

The graph below shows the results of the average energy consumption per message calculated at different spam rates. Please note that not only does the energy per message metric include the energy consumption of processing a message but also includes the consumption of issuing a message (computing PoW). The calculations were made by subtracting the reference and resting consumption from the measurement taken at a specific spam rate and then dividing by the spam rate. Tables 3 and 11 tabulate the data from Figure 3.

*Figure 10 - GoShimmer energy per message consumption at different spam rates*

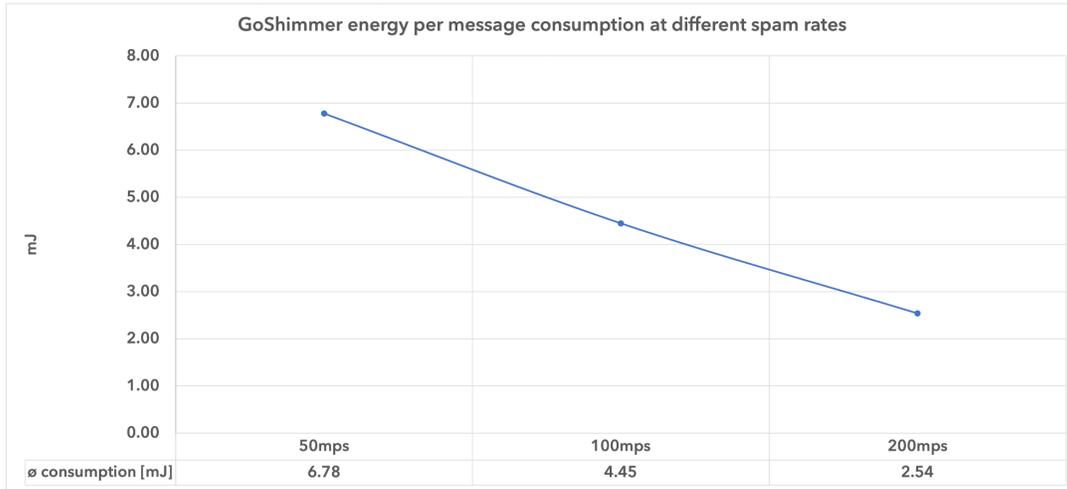

| Table 3 | 50mps | 100mps | 200mps |
|---|---|---|---|
| [mJ] | 6.78 | 4.45 | 2.54 |

The figure above shows that the energy per message consumption decreases while the spam rate increases. It is negatively correlated to the spam rate. The reason for this relationship will be explored in a future report.

## 4.5. Measurement results

In this section we provide details on the power consumption for each testing scenario. The power consumption is provided including the reference consumption.

### 4.5.1. Reference

Table 4 displays the reference measurements collected when the Raspberry Pi 4B is turned on without any GoShimmer nodes running, with power consumption displayed in watts.

| Table 4 | Total average | Minimum | Maximum |
|---|---|---|---|
| [W] | 2.648 | 2.494 | 2.914 |

| Standard deviation ± | Standard error for all three nodes ± | Power requirement compared to running a 60W light bulb |
|---|---|---|
| 0.107 | 0.015 | 4.41% |

### 4.5.2. Resting

Table 5 displays the Resting measurements collected when the GoShimmer private network runs with two to three activity messages per second (these are required for the network to function). Additionally, these results represent the minimum base power required for running a GoShimmer network.

| Table 5 | Per GoShimmer node (ø) | All three nodes (ø) | Minimum | Maximum |
|---|---|---|---|---|
| [W] | **0.9149** | 2.745 | 2.587 | 3.140 |

| Standard deviation ± | Standard error for all three nodes ± | Power requirement compared to running a 60W light bulb |
|---|---|---|
| 0.146 | 0.021 | 1.52% |

### 4.5.3. 50mps

Table 6 displays the measurements collected when the GoShimmer private network is spammed with 50mps.

| Table 6 | Per GoShimmer node (ø) | All three nodes (ø) | Minimum | Maximum |
|---|---|---|---|---|
| [W] | **1.254** | 3.761 | 3.279 | 4.312 |

| Standard deviation ± | Standard error for all three nodes ± | Power requirement compared to running a 60W light bulb |
|---|---|---|
| 0.250 | 0.035 | 2.09% |

### 4.5.4. 100mps

Table 7 displays the measurements collected when the GoShimmer private network is spammed with 100mps.

| Table 7 | Per GoShimmer node (ø) | All three nodes (ø) | Minimum | Maximum |
|---|---|---|---|---|
| [W] | **1.360** | 4.080 | 3.796 | 4.549 |

| Standard deviation ± | Standard error for all three nodes ± | Power requirement compared to running a 60W light bulb |
|---|---|---|
| 0.178 | 0.025 | 2.27% |

## 4.5.5. 200mps

Table 8 displays the measurements collected when the GoShimmer private network is spammed with 200mps.

| Table 8 | Per GoShimmer node (ø) | All three nodes (ø) | Minimum | Maximum |
|---|---|---|---|---|
| [W] | **1.423** | 4.268 | 4.101 | 4.445 |

| Standard deviation ± | Standard error for all three nodes ± | Power requirement compared to running a 60W light bulb |
|---|---|---|
| 0.070 | 0.014 | 2.37% |

Please note that the three nodes in the GoShimmer network running on the Raspberry Pi 4 were not able to consistently handle a 200mps rate. This means that the average presented in Table 8 does not accurately represent the power consumption at 200mps, but rather the average power consumption at a fluctuating spam rate at around 200mps. See the image below:

*Figure 11 - Visualization of the private GoShimmer network mps rate*

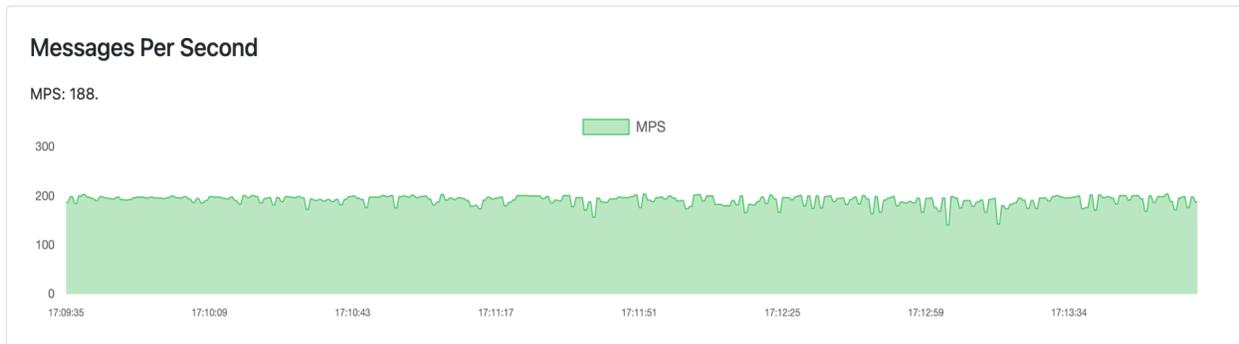

Even though the three-node docker private network is not able to handle a consistent 200mps spam rate, it does not confirm that IOTA 2.0 is not able to handle a 200mps network load. This is because the difficulties in processing at a constant rate of 200mps could have also been caused by the hardware limitations of the Raspberry Pi.

## 4.6. Data Normalization

To evaluate the relative difference in the power consumption at each spam rate, we subtract the reference average power consumption from each measurement. Results are displayed in milliWatts.

### 4.6.1. Scenario consumption per GoShimmer node - Reference

These values represent the power consumption per GoShimmer node of the different spam scenarios without the Reference (Table 4) consumption.

| Table 9 | Resting | 50mps | 100mps | 200mps |
|---|---|---|---|---|
| [mW] | 32.21 | 371.14 | 477.37 | 539.97 |

### 4.6.2. Scenario consumption per GoShimmer node - Resting and Reference

These values represent the power consumption of the different spam scenarios without the reference (Table 4) and resting (Table 5) consumption.

| Table 10 | 50mps | 100mps | 200mps |
|---|---|---|---|
| [mW] | 338.93 | 445.16 | 507.76 |

## 4.7. Energy consumption per message

The energy per message metric is mentioned in the previous Chrysalis benchmarking as well as in this report. When referring to the energy per message consumption, one is actually referring to a compound of processes. The equation below illustrates what the energy consumption per message means.

Let E(·) be the energy consumption of a certain process. Then the total energy consumption of a message can be understood as the following:

$$E_{(message\ added\ to\ the\ ledger)} =$$
$$E_{(issuance\ incl.\ PoW)}$$
$$+$$
$$(N-1) * E_{(processing\ message)}$$

For a message to be added to the ledger, a node has to issue the message and the other nodes present in the network have to process it. Issuing a message currently requires a small amount of PoW to be executed. The E(processing message) is dependent on the number of nodes in the network, minus the issuing node (thus we multiply by [N-1]). For example, if we have five nodes and one of them is the issuing node, then we have four nodes that are only processing a message and not computing PoW.

Due to the lack of a significant PoW requirement in GoShimmer, the processing and issuance of a message cannot be practically distinguished. This is because there is little difference in the processing of messages between issuing nodes and nodes that simply process messages. Thus, in our experiments we consider the energy consumption per message to be independent of whichever node issued a given message. This is different from the Chrysalis setup, where we clearly need to distinguish between issuing nodes (performing PoW) and receiving nodes.

### 4.7.1. Calculating the energy consumption per message

We assume that a transaction consumes energy inversely proportional to the spam rate. Thus we can calculate the energy required to process a single message. Below is an example equation for the spam rate of (x)mps:

$$E_{M,x} = (\,(P_{total,\,x} - P_{ref}) - (P_{rest} - P_{ref})\,) / x$$

$x$ = the spam rate (mps)

$E_{M,x}$ = the energy consumption required to process a single message at a spam rate of $x$ mps.

$P_{total,\,x}$ = total power consumption at x mps

$P_{ref}$ = reference power consumption

$P_{rest}$ = resting power consumption

In the equation above, we subtract the own consumption of the Raspberry Pi (reference) and the base consumption of running a node without spam (resting). The sum of both brackets is divided by $x$, as $x$ is the spam rate for which we are calculating the energy per message. Let us apply the above equation to calculate the energy consumption per message at 50mps spam rate of this report:

$$E_{M,50} = (\,(P_{tot,\,50} - P_{ref}) / 3 - (P_{rest} - P_{ref}) / 3\,) / 50 \text{ s}^{-1}$$

The reason why both inner brackets are divided by three is that the measurements for the total energy consumption at 50mps and the resting consumption were made for all three nodes of the private network. When applying the equation to measurements taken from a single node, one can use the first equation.

### Results

Table 11 shows the results when applying the above energy per message consumption equation to the total power consumption results of spamming at 50, 100 and 200mps.

| Table 11 | 50mps | 100mps | 200mps |
|---|---|---|---|
| Average energy/message (including issuance) [mJ] | 6.78 | 4.45 | 2.54 |

*See Figure 10 from p.19*

## 4.8. Relative Analysis and Comparison

The results for the **power** consumption per node and the **energy** consumption per message of the IOTA 2.0 prototype can now be compared to the previous Chrysalis energy consumption results.

### 4.8.1. Energy consumption per message - comparison

In this section we compare the energy consumption **per message**. The GoShimmer values include the energy consumption of issuing a message (computing PoW). Table 12 aggregates data from Table 11 and Table 15 (from the initial Chrysalis energy benchmarking).

| Table 12 | GoShimmer | | | Chrysalis without PoW | | Chrysalis with PoW |
|---|---|---|---|---|---|---|
| Spam rate | 50mps | 100mps | 200mps | 50mps | 100mps | 0.07mps |
| [mJ / message] | 6.78 | 4.45 | 2.54 | 1.21 | 1.19 | 4,026.44 |

| Table 13 - GoShimmer (**energy** / **message**) compared to | | |
|---|---|---|
| Processing a message | | Issuing a message |
| Chrysalis without PoW | | Chrysalis with PoW (compared to a 50mps scenario from GoShimmer) |
| 50mps | 100mps | 0.07tps |
| +459.01% | +274.21% | -99.83% |

The energy consumption of issuing a single message in GoShimmer compared to the mainnet parameters in Chrysalis, (i.e. the scenario with PoW), shows a significant reduction of 99.83%. This is mainly because the amount of PoW is very little in GoShimmer compared to the PoW in the Chrysalis network. As mentioned in the introduction, the reason why the PoW in the prototype is so low is to demonstrate how IOTA 2.0 would behave when the IOTA Congestion Control Algorithm (ICCA), which controls access, is fully implemented. In IOTA 2.0, PoW is replaced with Adaptive PoW[23]. This proposed mechanism makes the amount of PoW required to issue a message dependent on the number of messages the node wants to issue in a certain timeframe. In other words, if a node wants to issue a large number of messages in a short amount of time, it has to compute a lot more PoW than normally. The ICCA allows for nodes that do not spam the network to only have to compute a very small amount of PoW or none at all. This small amount is represented by the low PoW difficulty in this GoShimmer setup (set to 2), and so we observe a drastic reduction in energy consumption when comparing GoShimmer to Chrysalis. We note that the official GoShimmer DevNet PoW difficulty is (at the time of writing) set to a higher difficulty (21) to temporarily substitute the access control algorithm. The variation in the PoW difficulty for different networks even in the prototype means that the overall energy consumption measurements might be subject to change when the above-mentioned algorithms are fully implemented in the future. In summary, when comparing GoShimmer to Chrysalis, the reduction in energy consumption is dominated by the reduced consumption per issued message.

Finally, as one can see for 50mps and 100mps, the energy consumption at a given node for simply processing a message in GoShimmer is higher than the energy consumption of a message in Chrysalis without PoW. This was expected since the consensus-related modules as well as potentially other components can be more demanding in GoShimmer than in Chrysalis. As we only measured the energy consumption per message, we can also expect that future research which measures the energy consumption per transaction for GoShimmer will also show higher values when compared to Chrysalis.

Even though we observe a drastic reduction in energy consumption when comparing GoShimmer to Chrysalis with PoW, we must consider that this represents only a reduction in the energy required for issuing a message. We want to highlight that the PoW is not necessarily the main source of energy consumed for a given message. Consider a network of **N** nodes: When a node issues a message, the message has also to be processed **N-1** times and, thus, the total energy consumption of receiving nodes scales linearly with **N** while it is constant for the issuing node. As a consequence the energy consumption required for processing messages in larger networks can become dominant, i.e. become larger than the energy consumption required for the issuance of messages. Therefore the reduction in the energy consumption when comparing GoShimmer to Chrysalis, due to the lack of PoW, is particularly significant in smaller networks. In 2021, the IOTA mainnet underwent an upgrade called "Chrysalis", which brought about a significant energy reduction of 71.61% for the processing of received transactions at a network load of 50tps[24]. With this upgrade, the consumption for message processing is less than in Goshimmer (see Table 13). As a consequence, for **very large** networks and the current PoW difficulty, the energy consumption in Chrysalis can be less than in GoShimmer. However, the prototype code has yet to be improved for efficiency. In addition, it is possible that the PoW difficulty in Chrysalis might have to be increased if the network would be subject to spam attacks. As such the IOTA 2.0 solution has additional room for energy efficiency improvements.

### 4.8.2. Power consumption per node - comparison

In this section we compare the power consumption per node. Below we show data from Table 9 compared to Table 11 of the initial Chrysalis energy consumption benchmarking.

---

[23] https://v2.iota.org/how-it-works/module6a
[24] https://blog.iota.org/internal-energy-benchmarks-for-iota/

| Table 14 | Chrysalis | GoShimmer |
|---|---|---|
| mps / tps rate | Resting (0tps) | Resting (base activity) |
| [mW] per node | 15.67 | 32.21 |

| Table 15 - Chrysalis to GoShimmer comparison (**power per node without base consumption of device**) ||
|---|---|
| Resting (0tps) | Resting (base activity) |
| Raspberry Pi 4 (4 cores) | Raspberry Pi 4B Rev 1.1 |
| +105.57% ||

The overall power consumption of running a node shows an increase in comparison to the Chrysalis power consumption. As aforementioned, this is possibly due to novel IOTA 2.0 modules, whose performance is not yet optimized. Previously, consensus in the Chrysalis network was achieved when a message references a milestone message from the Coordinator. With the Coordinator removed in GoShimmer, all nodes now play a role in the consensus process. At the core of the new consensus mechanism is a voting protocol that considers the opinions of all other nodes in order to confirm messages and decide where and how to attach new messages to the Tangle[25]. This may result in a higher load on the individual nodes in comparison to the Chrysalis network. The additional demand on the nodes is around 105.57%. The comparison in Table 15 should thus be seen as an analysis of what the IOTA 2.0 modules now additionally demand of an individual node in comparison to the Chrysalis network. However, this increase in power consumption per node is not dramatic. As mentioned previously, the majority of the energy consumption in the IOTA network is a result of the processing of messages.

# 5. Model for estimating the total annual energy consumption of an IOTA network

## 5.1. Background

One should note that the metrics presented in this report (such as the energy per message consumption) should not be solely used to draw conclusions about a protocol's overall energy consumption. This is because the energy consumption per message only represents one of several contributing factors (another is the power consumption per node). An example that highlights this is Solana's energy consumption per message as reported by the Crypto

---

[25] https://v2.iota.org/how-it-works/introduction

Carbon Ratings Institute[26]. When the energy consumption of Algorand, Avalanche, Cardano, Polkadot and Solana were compared, Solana's energy per message consumption was reported to be the lowest value compared to the other protocols above. However, Solana's overall yearly electricity consumption ranks as the highest among them.

To accurately compare versions of the IOTA protocol's energy consumption with other networks, we have to derive an equation to calculate the total annual energy consumption of the network. Moreover, having an estimated annual energy consumption would allow us to also calculate an estimated cost of running the network as well as estimating the network's carbon footprint in the future.

The Digiconomist Website, created by Alex de Vries, provides a model to estimate the Bitcoin network's annualized energy consumption.[27] The model for estimating the annual energy consumption includes the total mining revenues made by Bitcoin validators. We cannot use a similar approach for estimating the annual energy consumption of an IOTA network. This is because running a node has no direct economic incentives (an IOTA network is feeless with no miners paid to validate transactions), which requires a different method to determine the total energy used. Thus, a new model for IOTA's network consumption has to be derived. The following pages will present this approach.

## 5.2. The model

To define an equation for the complete energy consumption of the IOTA network, we need to consider all contributing factors. The most apparent factor, which is also used in public comparisons, is the annual energy consumption of a node. We also discuss the energy consumption of a single message, more precisely its total energy consumption across the network. Considering the power consumption of a node, one can approximate the total energy consumption, which is the sum of every individual node running on different hardware types.

> Example: We have a total of 10 nodes in the network. Two nodes run on hardware A and eight nodes run on hardware B. The total energy consumption of all nodes is thus the energy consumption of node A * 2, added to the energy consumption of node B * 8 in this example.

We can thus write the above statements into this notation:

| $$\sum_{i \in hardware}^{H} P_{base,i} \cdot N_i$$ | Where **P(base, *i*)** equals the power consumption of a node running on hardware *i* at base activity and **$N_i$** equals the number of nodes running on hardware *i*. **H** is the number of hardware types present in the network. |
|---|---|

To obtain the total annual energy consumption, we multiply the above power consumption by the number of seconds in a year.

$$\sum_{i \in hardware}^{H} P_{base,i} \cdot N_i \cdot 86400\ s \cdot 365$$

Additionally, we have to consider that not every node is a Raspberry Pi running in a home environment. Most nodes are run on the virtual private servers of privately-owned cloud providers. This is significant because there is an additional overhead to the energy consumption of running a node on a data center, which is the additional energy required to run cooling, air conditioning, lighting and other IT infrastructure necessary for maintenance. The Power Usage Effectiveness (PUE) metric is the most popular method of calculating the energy efficiency of data centers. It is defined as the ratio between the total amount of energy used by a data center and the energy delivered to the IT equipment (servers). We can use this ratio as a factor and multiply it by the energy consumption of a node to estimate the total infrastructure energy consumption for a year. We define the first part of the model as E(base).

$$E_{base} = \sum_{i \in hardware}^{H} PUE_i \cdot P_{base,i} \cdot N_i \cdot 86400\ s \cdot 365$$

Next we have to incorporate the energy consumption of processing messages into our equation. Because the energy consumption per message changes based on the spam rate at which the measurements are being made, we have to incorporate the spam rate into our calculations. The most accurate calculation would be to account for the variability in the rate and integrate the energy consumption over a year. However, for the sake of simplicity, we assume that the rate remains constant for at least a day. Thus for a given day *d* we have an average spam rate $x_d$. The power consumption of a device of type *i* at a given rate *x* is then:

| $P_{M,i\ at\ x} = E_{M,\ i,\ at\ x} \cdot x$ | Where $E_{M,i}$ is the energy consumption of processing a single message on hardware *i* at spam rate *x*. |

The calculation above will give us the total power consumption of the messages processed in a second. To get an annual energy consumption, we will multiply the above equation by the number of seconds in a day (since for a given day *d* the rate is assumed fixed at $x_d$), and sum over each day of the considered year period.

$$E_{M,total,\ i} = \sum_{d=1}^{365} N_i \cdot E_{M,\ i,\ at\ x} \cdot x_d \cdot 86400\ s$$

The last step is to add the ratio of the nodes running on hardware *i* to the total number of nodes in the network. This is because all nodes in a hardware group *i* do not process all of the messages on their own; the complete network processes all messages. We can represent this by the notation below. The total annual energy consumption required to process all messages in a year by all of the hardware groups is then:

$$E_{M,total} = \sum_{i \in hardware}^{H} PUE_i \cdot E_{M,total,\ i}$$

## 5.3. Best estimate "T.A.E.C." model for a data-only IOTA network

We now combine the above equations into a best estimate model for the total annual energy consumption of an IOTA network only processing data messages. The total energy consumption annually is thus E(base) added to E(M,total).

$$E_{annually} = E_{base} + E_{M,total}$$

$E_{annually}$
Is the total average energy consumption in WattSecond (or joules) of the network for a year.

$i$
Represents nodes running on specific hardware running on a specific data center. If the GoShimmer network only had nodes running on Raspberry Pi 4s and Raspberry Pi 3s then we would repeat the expressions for $E_{base}$ two times with the appropriate variables (H=2).

$PUE_i$
The ratio between the total amount of energy used by a data center and energy delivered to the server on which hardware *i* is running. This factor is the additional energy needed for cooling, lighting and other infrastructure in a data center.

$P_{base,i}$
The power consumption of a node running on hardware *i* at base activity.

$N_i$
The number of nodes running on hardware *i* in the network.

$N_{total}$
The total amount of nodes running in the network.

$E_{M,total,i}$
Is the total amount of energy spent by a group of nodes running on hardware *i* by processing messages for a year.

$E_{M,total}$
Is the total amount of energy spent by all nodes in the network to process all of the messages for a year.

$E_{base}$
Is the total energy consumption of running all nodes of different hardware types in the network for a year.

$x_d$
The average mps rate of day *d.*

$E_{M, i, at\ x}$
Energy consumption per message on hardware *i* at spam rate $x$.

To make the equations less abstract, they will be applied in an example. See the next page for the calculations and explanations.

## 5.4. Example - hypothetical network calculation

To showcase how this model is used to derive a total annual consumption figure, the following paragraphs present a calculation of the total annual energy consumption of a hypothetical GoShimmer network where all of its nodes are running on Raspberry Pi 4Bs. This example uses this report's data about GoShimmer energy consumption as well as external analytics. Additionally, the table below presents the specific data used in the calculation. Please note that the model considers that each node only runs on one hardware device.

| Type | Value | Explanation |
| --- | --- | --- |
| $PUE$ | 1.59 | For the sake of explaining the calculation model and to make a conservative estimate that accounts for potentially omitted energy consumption, we will use the global average PUE value of 2021 as reported by the Uptime Institute[28]. In an ideal scenario, we would need the PUE value of every data center on which IOTA nodes are running. However, we do not have this data yet. |
| $N_{total}$ | 450 | For this example we chose the total number of nodes currently publicly available through auto peering in the Chrysalis network (current IOTA mainnet). Data from: Thetangle.org. Retrieved April 12, 2022, from https://thetangle.org/nodes |
| $N_1$ | 450 | As our hypothetical network only has nodes running on Raspberry Pi 4Bs, N = 450 and H = 1, as we only have to repeat measurements for one hardware type. |
| $P_{base,1}$ | 2.680131 [W] | This value is the reference consumption added with the power consumption per node. Tables 4 and 9 tabulate the data. |
| $E_{M, 1, at\ x}$ | 6.78 [mJ] | Taken from the results of Table 11 |
| $x_d$ | 50 [mps] | To simplify this calculation example, we will set the spam rate of 50mps constant for every day. |

**Applying the equations**

When we insert the values from the above table, the estimated total annual energy consumption of our hypothetical network is at **68,123,687,160.320** [Ws] for one year. See below for conversion to kWh and TWh.

| kWh for one year | TWh for one year |
| --- | --- |
| 18923.25 | 0.000018923 |

---

[28] Bizo, D., Ascierto, R., Lawrence, A., & Davis, J. (2021, September 1). *Uptime Institute Global Data Center Survey 2021*. Uptime Institute. Retrieved March 16, 2022, from https://uptimeinstitute.com/

In the introduction, we mentioned that Bitcoin consumes an estimated 204.5 TWh annually. The table below highlights the difference between Bitcoin and IOTA.

| Bitcoin estimated total annual power consumption | Percentage of the hypothetical IOTA network consumption in comparison to Bitcoin's consumption |
|---|---|
| 204.50 [TWh] | **0.000009%** |

This comparison shows that a hypothetical IOTA network running on 450 Raspberry Pi 4Bs with a PUE value of 1.59 (which is usually only applicable for nodes running on data centers) and running at a constant rate of 50mps daily would only consume an estimated **0.000009%** of the currently-estimated annual Bitcoin network energy consumption. Additionally, having a kWh value for our hypothetical IOTA network consumption allows us to compare it with a more tangible metric, such as the per capita energy consumption (this metric includes electricity, but also other areas of consumption.)[29]. The graph below shows that the IOTA network would only require **43.30%** annually of the average annual energy consumption of a person in Germany (in 2019). Even though we used parameters that are similar to the actual IOTA mainnet, in reality, the IOTA 2.0 mainnet will probably consume more. This is because most nodes are run on hardware that consume a lot more energy than what a Raspberry Pi consumes. Thus, the above comparison should only be interpreted as a way to give an impression of which energy consumption scale IOTA is operating at.

*Figure 12 - Hypothetical IOTA network consumption compared to the average per capita consumption in Germany*

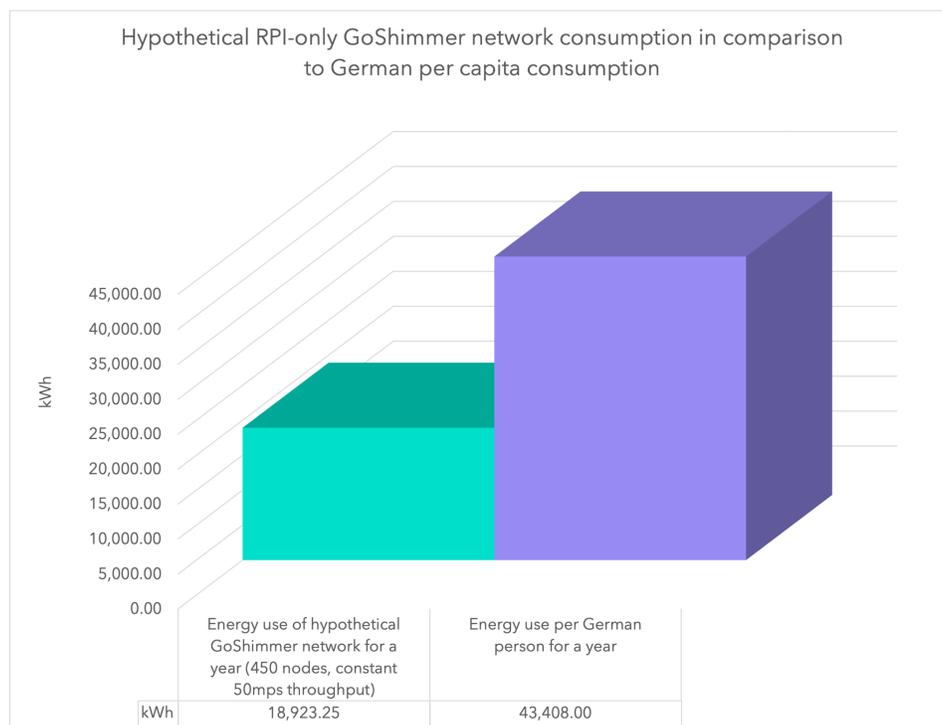

---

[29] *Statistical Review of World Energy 2021, all data 1965-2020.* BP Global. (2021, July). Retrieved March 25, 2022, from https://www.bp.com/en/global/corporate/energy-economics/statistical-review-of-world-energy.html

We note that, because the total throughput is limited, our model shows an almost linear dependency of the total energy consumption on the number of devices. The predicted low energy consumption in the above example, which is comparatively even less than that of German per capita consumption, indicates that, even when scaling the number of devices by a few magnitudes, the solution continues to provide an energy-efficient solution for a DLT.
On the other hand, it should also be considered that, as the throughput increases with the number of devices, additional scaling solutions must be considered. However, a scaling solution would impose non-linear effects on the energy consumption, a discussion of which is outside the scope of this report.

Please note that the energy use per person in Germany for a year represents not only the electrical energy consumption but additionally the energy used for transport, heating and cooking.

## 5.5. Applying the model to Chrysalis

With a model defined for an annual energy consumption figure for an IOTA network, we can also apply the data from the April 2021 benchmarking to calculate the total energy consumption of a hypothetical Chrysalis network. As previously mentioned, comparing only individual metrics between networks can result in misleading conclusions about a protocol's overall energy consumption. Thus, we compare the annual energy consumption of the two hypothetical GoShimmer and Chrysalis networks. When we input 450 nodes, a constant network activity of 50tps, a PUE of 1.59 and the energy consumption metrics from the Chrysalis benchmarking, the total energy consumption of this hypothetical Chrysalis network is at 20,742.95 kWh for one year. This comparison confirms that both GoShimmer and Chrysalis consume **similar** amounts of energy at the above mentioned input parameters.

*Figure 12 - Hypothetical IOTA network consumption compared to average German person's consumption*

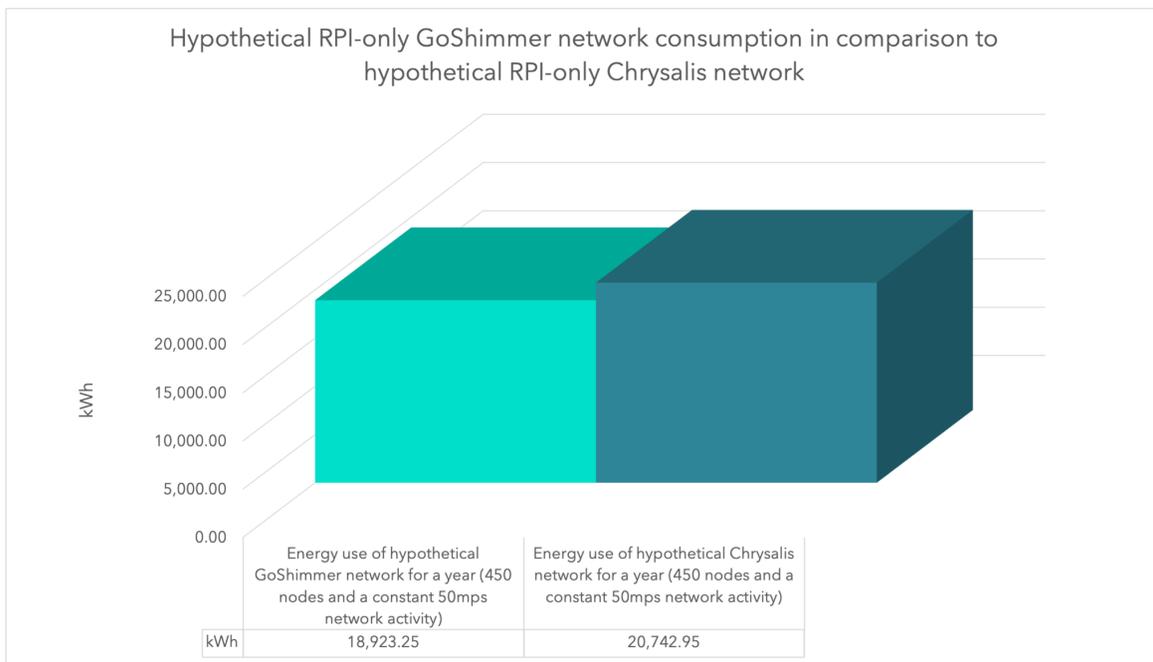

The effect of a reduction in only the issuance of messages due to a lack of PoW in GoShimmer (mentioned in section 4.8.1) becomes apparent in this example comparison. Even though we observe drastic reductions in the energy consumption per message in GoShimmer compared to Chrysalis with PoW, this reduction is only for the issuance of messages. Thus, in larger networks with higher network activity, the higher energy consumption for processing messages in GoShimmer becomes dominant and results in only the observed **8.77%** reduction to the overall energy consumption in the example used.

# 6. Conclusion

Overall, from the results presented in this report, it can be concluded that GoShimmer 0.8.3 results in energy consumption optimizations in comparison to Chrysalis for the issuance of messages. The improvement is mainly due to the removal or reduction of PoW requirements for nodes when issuing messages in GoShimmer, while in Chrysalis there remains a higher PoW score. The reason for the reduction of the PoW requirement in GoShimmer is to demonstrate how IOTA 2.0 would behave when the IOTA Congestion Control Algorithm, which controls access, is fully implemented. In IOTA 2.0, PoW will be replaced by other algorithms such as Adaptive PoW[30] and Mana[31], which allows for nodes that do not spam the network to only have to compute a very small amount of PoW.

However, when solely looking at the energy consumption of messages in Chrysalis without PoW, one can see that the GoShimmer energy per message consumption is higher than in Chrysalis. The cause of an increase in consumption can be new components with new functionalities, as well a missing optimisation of these components and functionalities. This increase in consumption for processing messages and a reduction in the consumption for issuing a message results in the effect that smaller GoShimmer networks will show an overall improvement to the total annual energy consumption. In larger networks, however, as the number of processed messages increases linearly with the number of nodes, we observe that the higher energy consumption for processing a message in GoShimmer becomes dominant and the reduction in energy consumption for issuing a message becomes insignificant. This means that large Chrysalis networks will probably consume less energy than a GoShimmer network of the same size. As GoShimmer is a research prototype node implementation, future optimizations to the code can bring improvements to the energy consumption for processing a message.

Consensus in the Chrysalis network was previously achieved when a message was referenced by a milestone message from the Coordinator. With the Coordinator removed in GoShimmer, all nodes now play a role in the consensus process[32]. The increased involvement of all nodes in network consensus can result in a higher load on the individual nodes in comparison to the Chrysalis network.

Another observation that can be made from the measurement results is that running GoShimmer at base activity (2-3mps activity messages) requires a low amount of energy. The majority of the energy consumption in spam scenarios comes from the additional processing of the spammed messages. However, it can be observed that the energy consumption per message decreases when the spam rate increases.

The power consumption per node and per message metrics are useful for developing estimates of consumptions. However, they are not suited for individual comparisons to other protocols. For this purpose, a model for estimating the annualized total energy consumption of an IOTA network was defined. The results of applying the model to the available data shows that a hypothetical GoShimmer network of 450 nodes with a constant network load of 50mps would require **0.000009%** of the estimated annual energy consumption of the Bitcoin network to function. Moreover, the same hypothetical GoShimmer network would require **43.30%** of the average annual per capita energy consumption in Germany in 2019. Having a model for an annual power consumption figure is significant, as it can inform about a protocol's environmental impact and serves as a tool to compare energy consumption profiles of different networks. It also allows us to make more informed decisions on how to improve the performance. Additionally, having a total figure increases transparency, which in turn allows for regulators, institutions and investors to make more educated decisions in the DLT sector.

---

[30] https://v2.iota.org/how-it-works/module6a
[31] https://blog.iota.org/explaining-mana-in-iota-6f636690b916/
[32] https://v2.iota.org/how-it-works/introduction

# 7. Limitations

The objective of this report is to evaluate the energy consumption profile of the IOTA 2.0 prototype and to make it tangible to the average reader. However, the precision of values presented in this report and the comparisons made have limitations that should be considered.

## 7.1. Accuracy and hardware

A limitation of the measurements made for this report is their accuracy. The private GoShimmer network used for testing struggled to consistently handle a steady spam rate the higher the spam rate grew. The spam rate especially fluctuated when it approached 200mps. A possible explanation for the fluctuation could be the hardware limitations of the Raspberry Pi 4B used for testing. However, this would have to be confirmed in future research. Additionally, one should note that the values presented in this report are only applicable for nodes running on Raspberry Pi 4Bs. The energy consumption metrics will change for every different hardware type. Expanding on the hardware limitations: all of the hardware used in testing is subject to slight variations in performance and energy consumption because of manufacturing processes. An example of such variation is presented in a study by Yewan Wang et al., which observed a 7.8% variation in power consumption between 12 identical servers running the same tests[33]. Thus, slight differences in the averages presented in this report and future replicated tests can be expected. Furthermore, the ambient and internal temperature of the device measured also has a significant effect on the overall energy consumption, mainly through temperature-sensitive components such as the CPU.

## 7.2. Chrysalis energy benchmarking

The comparisons made between the results of GoShimmer's energy consumption and the results of Chrysalis' energy consumption are also limited by the setup and methodology of the testing done for the Chrysalis benchmarking. In a blog post about the benchmarking, the author states that all measurements were collected over a period of 10 minutes. This means that the values in the Chrysalis benchmarking have a lower accuracy than the measurements presented in this report. The consequence is therefore that the comparison made between GoShimmer and Chrysalis is only as strong as the measurement accuracy for the Chrysalis tests and the GoShimmer tests. Future research that replicates the Chrysalis setup might change the percentage differences presented in the comparisons of this report.

---

[33] Wang, Y., et. al. *Potential effects on server power metering and modeling*. CloudComp (2018) : 8th EAI International Conference on Cloud Computing, Sep 2018, Exeter, United Kingdom. pp.1-12, ff10.1007/s11276-018-1882-1ff. Ffhal-01869705f, from https://hal.inria.fr/hal-01869705/file/cloudcomp2018.pdf

## 7.3. Annual power consumption model

This report defines and presents a model for estimating the total annual power consumption of an IOTA network. However, the reader should be aware that every model is a simplified representation of reality built upon assumptions, which themselves have limitations. One of them would be that the model might exclude currently-unknown contributing factors to the overall energy consumption of the network. Currently unknown factors could be omitted from our results or assumptions, as we are, for example, only testing a network with three nodes. A possibility could be that future research focused on a large number of nodes present in an IOTA network could identify additional behavior that justifies adding another variable to the model. In addition, the model's variables have limitations. One of them is the PUE variable. The PUE metric is mostly reported by the companies running the data centers, which results in a bias for the companies to skew the measurements and data collected for the PUE metric in such a way that it results in a lower value. This is because the PUE metric has also become a marketing tool for data center providers. Similar to the annual power consumption model, the PUE metric is also only as accurate as the percentage of contributing factors covered from the complete number of contributing factors. For example, excluding the inefficiencies of the power delivery system from the PUE calculation simply because of human error can cause an inaccurate PUE value.[34]

Another variable in the model that is subject to a limitation is $E(M,total,i)$. The main limitation for this variable is that we are using a daily average mps rate to calculate the total energy consumption per message for that day. Alternatively, we could repeat the $E(M,total,i)$ calculation for every second of the year instead of every day of the year to improve the precision of the results or average over periods of differing mps rates.

Moreover, the model does not consider the processing of value transactions. Only the energy consumption of processing data messages is represented in the equations. This is a limitation, as there might be a difference in the energy consumption between processing data messages and value transactions in GoShimmer. At the time when the measurements were conducted and the report was written, tooling for spamming value transactions was not readily available. Future research can explore the energy consumption difference between messages and transactions and add another variable to the model if a significant difference between the two types of transactions (data and value) exist.

# 8. Outlook and future research

As the IOTA technology advances, so should research into its energy consumption. This report should thus not be seen as a final destination for the energy consumption research for IOTA 2.0. Armed with a defined model for the total annual energy consumption of an IOTA network, we hope in the future to be able to estimate the annual power consumption of the IOTA mainnet, rather than estimating a hypothetical network with assumed values. In order to meet this objective we need to measure the energy consumption per node and per message for every hardware type on which IOTA nodes are running. Although this task is difficult, it can be simplified by only measuring the most significant hardware types. This means that future research can also focus on collecting data on network analytics to determine an estimate of the hardware distribution of the IOTA mainnet, i.e., study which nodes run on which devices. Moreover, analysis of GoShimmer and its energy consumption behavior will be used to continuously refine the annual power consumption model and to publish updates of the energy consumption figures presented in this report.

---

[34] Brady, G.A., Kapur, N., Summers, J.L., & Thompson, H.M. (2015). *A Case Study and Critical Assessment in Calculating Power Utilisation Effectiveness for a Data Centre*, from
https://www.semanticscholar.org/paper/A-Case-Study-and-Critical-Assessment-in-Calculating-Brady-Kapur/ab97e08ac9c82463a0f4ae6f1195d2e88c5b6409

IOTA    33

# 9. Acknowledgements

At the time of writing, all authors of this report were employed at the IOTA Foundation. The authors would like to thank Gal Rogozinski, Andrea Villa, Piotr Macek and David Phillips for reviewing content, helping to troubleshoot, assisting in the set-up and answering questions related to the prototype.